\documentclass[aps,prl,reprint,superscriptaddress]{revtex4-1}

\usepackage{graphicx}
\usepackage{lineno}
\usepackage{float}
\begin{document}
\preprint{}

\title{ \bf Measurement of $W^{\pm}$ single spin asymmetries and $W$ cross section ratio in polarized $p+p$ collisions at $\sqrt{s}=510$ GeV at STAR}


\author{Devika Gunarathne}
\email[]{devika@temple.edu}
\affiliation{Temple University}

\collaboration{for the STAR collaboration}

\date{\today}

\begin{abstract}
We present the preliminary results of measurements of single spin asymmetries, $A_L$ for $W^\pm$ boson production in longitudinally polarized p+p collisions at $\sqrt{s}=510$ GeV and measurements of cross section ratios, $\sigma _{W^+} / \sigma _{W^-}$, for $W^+,W^-$ boson production in $p+p$ collisions at $\sqrt{s}$ = 500 and 510 GeV. The asymmetry measurements are based on 246.2 pb$^{-1}$ of data taken in the RHIC 2013 run and the cross section ratio measurements are based on 102 pb$^{-1}$ of data taken during RHIC 2011 and 2012 runs by the STAR experiment. While the asymmetry results are shown as a function of the decay lepton  pseudorapidity, $\eta_e$, the cross section results are shown as a function of both $\eta_e$ and W boson rapidity, $y_W$ in the mid rapidity region ($|\eta_e|<1$). At these kinematics, $W^\pm$ single spin asymmetries provide a theoretically clean probe of the proton's polarized quark and antiquark distributions and the W cross section ratio provides sensitivity to unpolarized sea quark distributions at the scale of the W mass. The asymmetry results are consistent with recently published STAR $A^{W^\pm}_L$  results based on the data collected during RHIC 2011 and 2012 runs which showed a preference for a sizable, positive up antiquark polarization in the range $0.05 < x < 0.2$. The new preliminary results can be considered as the most precise results of $A^{W^\pm}_L$ in the world to date, with uncertainties reduced by 40\% in comparison to the published results.
\end{abstract}


\maketitle

\section{I. INTRODUCTION} \label{sec:intro}
There has been steady progress over the past few decades in terms of understanding the spin structure of the nucleon, one of the fundamental questions in nuclear physics. In the relativistic quark parton model, the spin of the proton was naively explained~\cite{dis:rqpModel} by the alignment of spins of the valence quarks. However, as of our current knowledge~\cite{dis:EMU}, the valence quarks, sea quarks, gluons and their possible orbital angular momenta are all expected to contribute to the overall spin of the proton. Despite this significant progress, the individual polarizations of quarks and antiquarks are yet to be understood precisely.\\
\indent
According to the spin sum rule introduced by Jaffe and Monahar~\cite{dis:sumrule} in 1990, the spin of the proton can be written in terms of its contributions from the intrinsic quark and antiquark polarizations, intrinsic gluon polarization and their orbital angular momenta. Polarized inclusive deep-inelastic scattering (DIS) experiments were able to strongly constrain the total quark contribution to the proton spin~\cite{dis:global}. However, DIS experiments were not sensitive to the flavor separated individual quark spin contributions. These were then measured by polarized semi inclusive DIS experiments (SIDIS), where a specific hadron is tagged in the final state. The helicity-dependent parton distribution functions (PDF)~\cite{dis:global} were extracted from global analyses using the world data of both DIS and SIDIS. Relatively large uncertainties remain in the polarized antiquark PDFs in comparison to quark PDFs mainly due to the large uncertainties found in the fragmentation functions~\cite{dis:frag} which were used in the global analysis. Over the years however, progressively more precise polarized SIDIS data, covering an enhanced kinematic range has become available~\cite{dis:newdata1, dis:newdata2, dis:newdata3}. Moreover the knowledge of the fragmentation process has increased, leading to the extraction of rather precise fragmentation functions~\cite{dis:FFs}. Furthermore, the global fitting tools used in various global analyses has improved over the years. Despite this significant progress, the current knowledge of antiquark helicity PDFs is still less precise in comparison to the valence sector~\cite{dis:lss10}. \\
\indent
The production of $W^\pm$ bosons in longitudinally polarized $p+p$ collisions at RHIC provides a unique and powerful tool to probe the individual helicity PDFs of light quarks and anti quarks at much larger $Q^2$ scale ($\sim$ 6400 $GeV^2$) set by the W mass. Due to the maximal  parity violating nature of the weak interaction, $W^{-(+)}$ bosons couple to the left handed quarks and right-handed anti quarks and hence offer direct probes of their respective helicity distributions in the nucleon. These distributions can be extracted by measuring the parity violating $A_L$, as a function of the decay electron (positron) pseudo rapidity, $\eta_e$. The longitudinal single-spin asymmetry is defined as $A_L =(\sigma_+-\sigma_- )/ (\sigma_++\sigma_-)$, where $\sigma_{+(-)}$ is the cross section when the  helicity of the polarized proton beam is positive (negative). At leading order, $W^+$ $A_L$ is directly related to polarized anti d and u quark distributions ($\Delta \bar d$, $\Delta u$) while  $W^-$ $A_L$ is directly related to polarized anti u and d  quark distributions ($\Delta \bar u$, $\Delta d$)~\cite{dis:rhicW}. \\
\indent
Considering the SU(3) flavor symmetry, since the mass difference of up and down quarks is small, equal numbers of up and down quark-antiquark pairs are expected to be produced perturbatively in the nucleon sea from gluon splitting. However in 1970's the first indication of up-down asymmetric sea came through early SLAC data suggesting the violation of the Gottfried Sum Rule (GSR)~\cite{dis:GSR}. Later on more concrete evidence supported the up-down asymmetric sea with surprising results from the FNAL E866 Drell-Yan (DY) Experiment~\cite{dis:E866}. The E866 data clearly showed that $\bar d \neq \bar u$, suggesting a non perturbative origin of the nucleon sea. However, theoretical calculations failed to explain the $\bar d / \bar u$ behavior at higher Bjorken-x values~\cite{dis:FAtheory}. Moreover, the most recent preliminary results of the SeaQuest E906~\cite{dis:seaquest} experiment where DY measurements have extended to larger $x$ values deviates from E866 results at higher $x$ values. \\
\indent
At leading order, the ratio of W cross sections, $\sigma^+_W / \sigma^-_W$ in $p+p$ collisions provides a direct measurement of the unpolarized flavor structure of the sea as shown in equation~\ref{eq:xsection} which is expressed in terms the unpolarized PDFs of up, down quarks and antiquarks.\begin{equation} 
\frac{\sigma ^+_W} {\sigma ^-_W} = \frac{u(x_1)\bar{d}(x_2) + \bar d(x_1)u(x_2)}{\bar u(x_1)d(x_2) + d(x_1)\bar{u}(x_2)}
\label{eq:xsection}
\end{equation}
The RHIC kinematic range is highly sensitive to the particular $x$ region where the two DY results disagree and the steadily increasing behavior of $\bar d / \bar u$ changes~\cite{dis:FAtheory}. Hence, measurements of W cross section ratio at RHIC provide an important and completely independent cross check of the up-down flavor asymmetry of the sea at much larger $Q^2$ values than the DY measurements.\\
\indent
This paper is organized as follows: section II provides a brief overview of the experimental aspects focusing on the use of various detector elements at STAR in terms of reconstructing and extracting W signal spectra from the data set. This section also explains the estimation and subtraction of the background from the W signal spectra. In section III we discuss the calculation of the W single spin asymmetry and W cross section ratio and present the preliminary results comparing to several theoretical calculations. Finally the last section provides a summary and outlook.
\section{II. ANALYSIS}\label{ana}
The data analyzed here for W $A_L$, were collected by the STAR experiment in RHIC 2013 running of longitudinally polarized $p+p$ collisions at $\sqrt{s}=510$ GeV. The total integrated luminosity of the data are 246.2 pb$^{-1}$, with an average beam polarization of 54\%. The data used for the W cross section analysis is the combination of data collected by the STAR experiment in RHIC 2011 and 2012 running of $p+p$ collisions at $\sqrt{s}$ = 500 and 510 GeV with a total integrated luminosity of 25 pb$^{-1}$ and 77 pb$^{-1}$ respectively. Both asymmetry and cross-section analyses whose final results are presented as a function of $\eta_e$, followed largely the same analysis procedure. The data were selected online using the same high energy trigger requirement in the calorimeter for both analyses. Next, similar steps for the reconstruction, extraction of the W signal, and the estimation and subtraction of backgrounds were followed. However, an additional step of correction for W detection efficiencies was involved for the cross section analysis. As for the cross-section analysis as a function of $y_W$, the details of the W reconstruction algorithm can be found in Ref.~\cite{run11paper}.\\
\indent
The STAR experiment~\cite{dis:starNIM} is well equipped to measure $A_L$ for $W^\pm$ boson production within a pseudorapidity range of $|\eta|<1$. $W^\pm$ bosons are detected via their $W^\pm \rightarrow e^\pm \nu$ decay channels. A subsystem of the STAR detector, the Time Projection Chamber (TPC) is used to measure the transverse momentum ($p_T$) of decay electrons and positrons and to separate their charge sign. Two other subsystems, Barrel and Endcap Electromagnetic Calorimeters (BEMC, EEMC) are used to measure the energy of decay leptons. A well developed algorithm~\cite{dis:run12paper} is used to identify and reconstruct $W^\pm$ candidate events by removing QCD type background events. In this algorithm, various cuts are applied at each level of the selection process based on the kinematics and topological differences between the electroweak process of interest and QCD processes. For example, tracks associated with $W^\pm$ candidate events can be identified as isolated tracks in the TPC that point to an isolated EMC cluster in the calorimeter. However, QCD type events have several TPC tracks point to several EMC clusters. In contrast to QCD background events, a large missing transverse energy opposite in azimuthal direction ($\phi$) can be observed for $W^\pm\rightarrow e^\pm \nu$ candidate events, due to the undetected neutrinos in the final state. This leads to a large imbalance in the vector $p_T$ sum of all reconstructed final-state objects in W candidate events, which is expressed as $\vec{p}^{~balance}_T$ in equation~\ref{eq:signPt1}. 
\begin{equation}
\vec{p}^{~balance}_T=\vec{p}^{~e}_T+\sum\limits_{\Delta R > 0.7} \vec{p}^{~jets}_T
\label{eq:signPt1}
\end{equation}
\begin{figure}[H]
\centering
\includegraphics[width=0.48\textwidth]{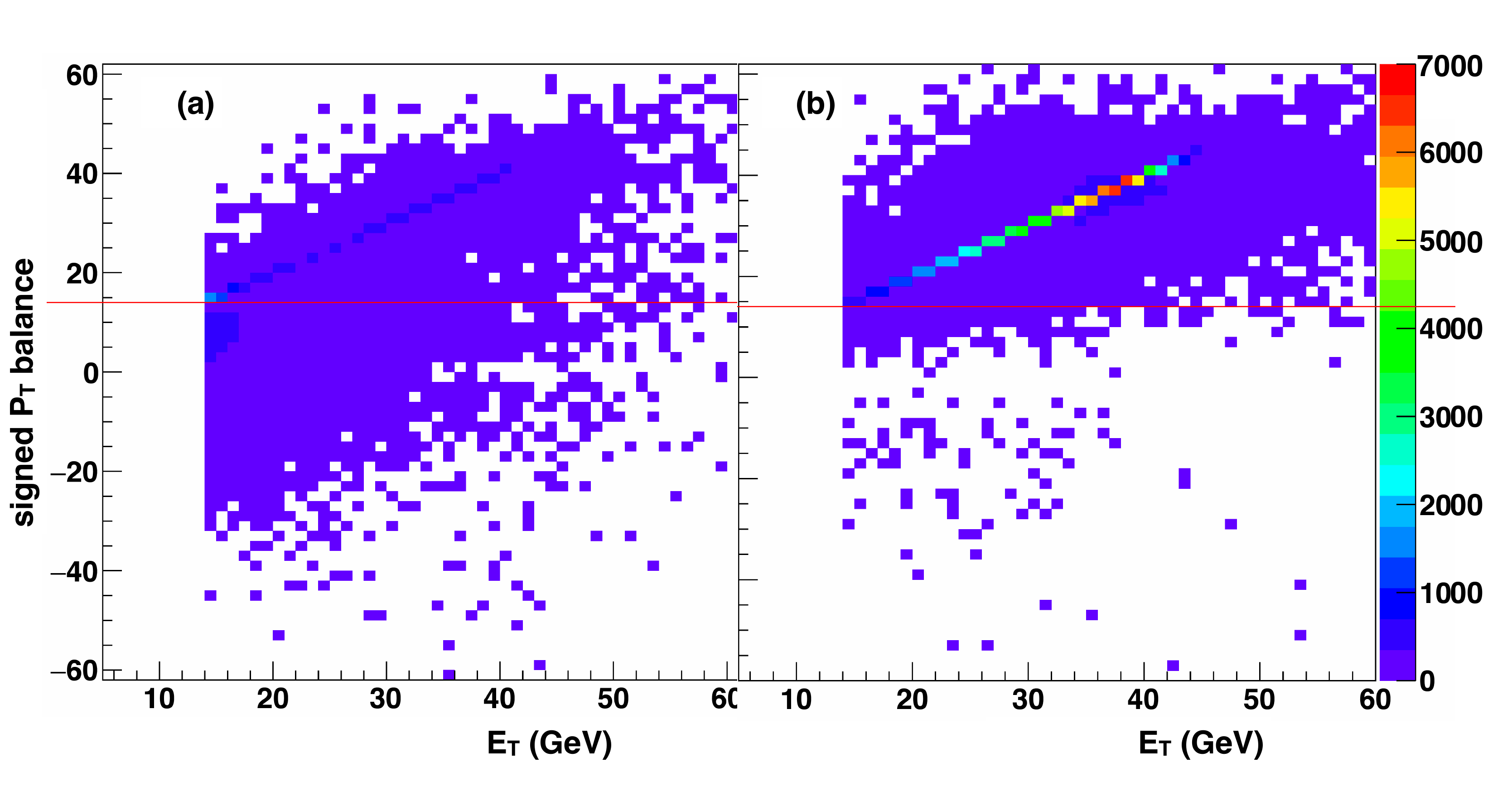}
\caption{Signed $p_T$-balance vs $E^e_T$ for data (a) and $W \rightarrow e\nu$ MC (b).}
\label{Fig:signPt}
\end{figure}
Here, the $\vec{p}^{~e}_T$ is the $p_T$ of the candidate lepton. The transverse momentum of jet like events, $\vec{p}^{~jets}_T$ is determined using the anti-$\it{k}_T$ algorithm~\cite{dis:antikt}, by reconstructing the $p_T$ from all reconstructed jets outside a cone of radius of 0.7 in $\eta-\phi$ space which is centered around the candidate lepton. A strong correlation is observed between $E^e_T$ and the scalar quantity, signed $p_T$ balance, which is defined in equation~\ref{eq:signPt2}. This can be seen clearly in Fig~\ref{Fig:signPt} (b) which shows the MC results simulating $W^\pm\rightarrow e\nu$ decays. \\
\begin{figure}[H]
\centering
\includegraphics[width=0.45\textwidth]{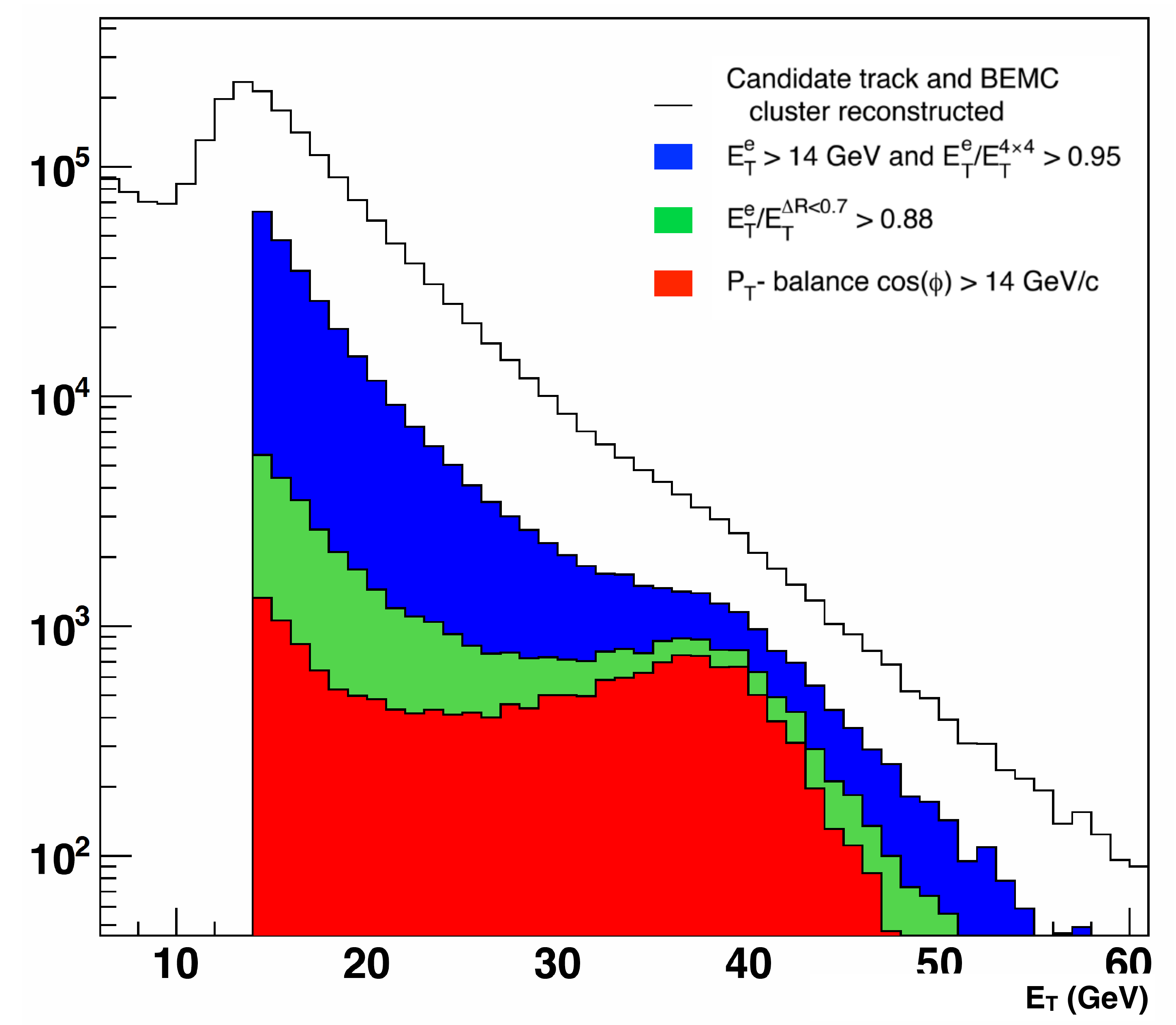}
\caption{Candidate $E^e_T$ distribution from the data after various selection cuts.}
\label{Fig:Et}
\end{figure}
\indent
After initially requiring reconstructed TPC tracks to have $p_T > 10$ GeV, candidate tracks were matched to a BEMC tower. The transverse energy, $E_T$ of each of four possible $2\times2$ clusters which contain the pointed track was computed and the cluster with largest summed $E_T$ was assigned as the candidate cluster. The energy corresponding to the candidate cluster was considered as the energy of the candidate, $E^e_T$ and was required to be greater than 14 GeV. Next, two isolation requirements were imposed. First, the ratio of $E^e_T$ to total energy of the $4\times4$ BEMC cluster surrounding the candidate cluster, was required to be grater than $95\%$. Second isolation requirement demanded that the ratio of $E^e_T$ to $E_T$ sum of the area within a cone of radius of 0.7 around the candidate track ($E^{\Delta R < 0.7}_T$) to be greater than $88\%$. As the final requirement, the signed-$p_T$ balance mentioned in equation~\ref{eq:signPt2} was required to be greater than 14 GeV, which is indicated by the red line in Fig.~\ref{Fig:signPt} (a). After all the selection cuts have been applied, the characteristic Jacobean peak in the $E_T$ distribution for mid-rapidity $W^\pm$ candidate events was observed near the half of the $W^\pm$ mass, as shown in Fig.~\ref{Fig:Et}.
\begin{equation}
signed\:{p_T}\:balance=\frac{(\vec{p}^{~e}_T).\vec{p}^{~balance}_T}{|\vec{p}^{~e}_T|}
\label{eq:signPt2}
\end{equation}
\begin{figure*}
\centering
\includegraphics[width=0.99\textwidth]{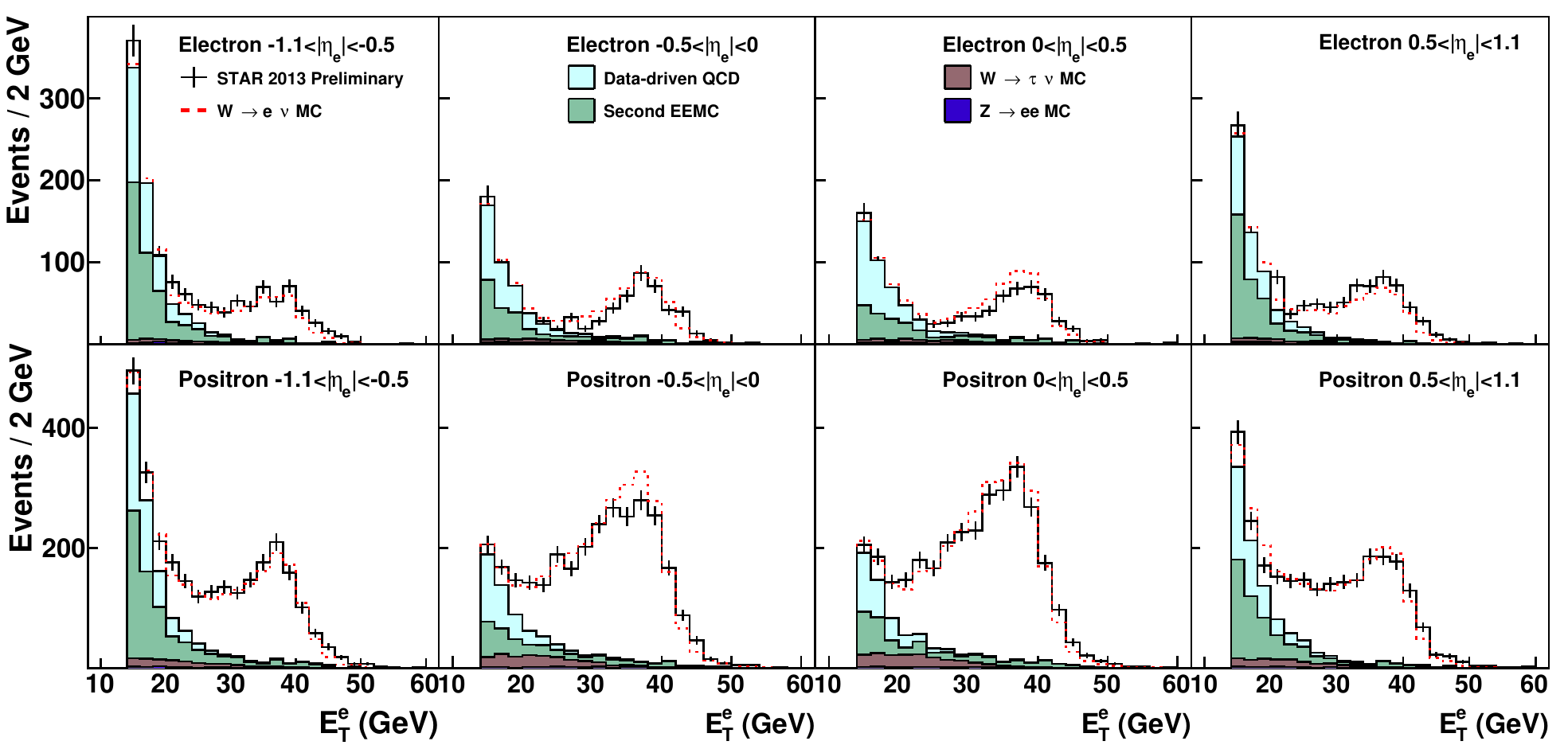}
\caption{$E^e_T$ distribution of $W^-$ (top) and $W^+$ (bottom) candidate events (black), various background contributions and sum of backgrounds and $W\rightarrow e\nu$ MC signal (red-dashed).}
\label{Fig:bg}
\end{figure*}
$W^\pm$ candidates were charge separated based on $e^\pm$ track curvature measured in the TPC. The charge separated $W^\pm$ yields as a function of $E^e_T$ are shown in Fig.~\ref{Fig:bg} for four different $\eta$ bins, along with the estimated residual background contributions from electroweak processes and QCD processes. 
The $W^\pm\rightarrow\tau^\pm\nu_\tau$ and $Z/{\gamma^*}\rightarrow e^+ e^-$ electroweak background contributions were estimated from Monte-Carlo (MC) samples.
The MC samples were simulated by generating the respective events using PYTHIA 6.422~\cite{dis:pythia} and passing through the STAR GEANT\cite{dis:geant} model and embedding into STAR zero-bias triggered events. In comparison to QCD, relatively small electroweak background contributions were estimated.\\
\indent
The selection process of W candidate events described above is designed to remove significant amount of QCD type background events. However a certain amount of QCD background will still be present in the signal region. This contribution originates primarily from events which satisfy candidate $W^\pm$ isolation cuts but contain jet fragments which escape the detection outside the STAR acceptance. Two procedures referred as ``Second EEMC" and ``Data-driven QCD", were used to estimated these backgrounds associated with the acceptance ranges $-2<\eta < -1.09$ and $|\eta|>2$ respectively. The backgrounds treated with the Second EEMC procedure refers to the $e^\pm$ candidate events that satisfy W isolation requirement which had an opposite-side jet fragment in the range $-2 < \eta < -1.09$, where an EEMC does not exist in opposite to the real EEMC ($1.09<\eta < 2.0$) at STAR. Therefore, this opposite-jet fragment had escaped the detection leading respective events to satisfy W candidate requirements. The magnitude of this background contribution was estimated using the real EEMC~\cite{dis:run9paper}. The backgrounds treated with Data-driven QCD procedure were referred to QCD background events that satisfied W candidate selection criteria due to the escape of detection of a jet fragment similar to the Second EEMC case, but in the range of $\eta < |2|$. This component of the background was estimated using a data-driven procedure~\cite{dis:run9paper} as a function of $E^{e}_{T}$.
\section{III. RESULTS}\label{results}
\begin{figure}[!h]
\centering
\includegraphics[width=0.48\textwidth]{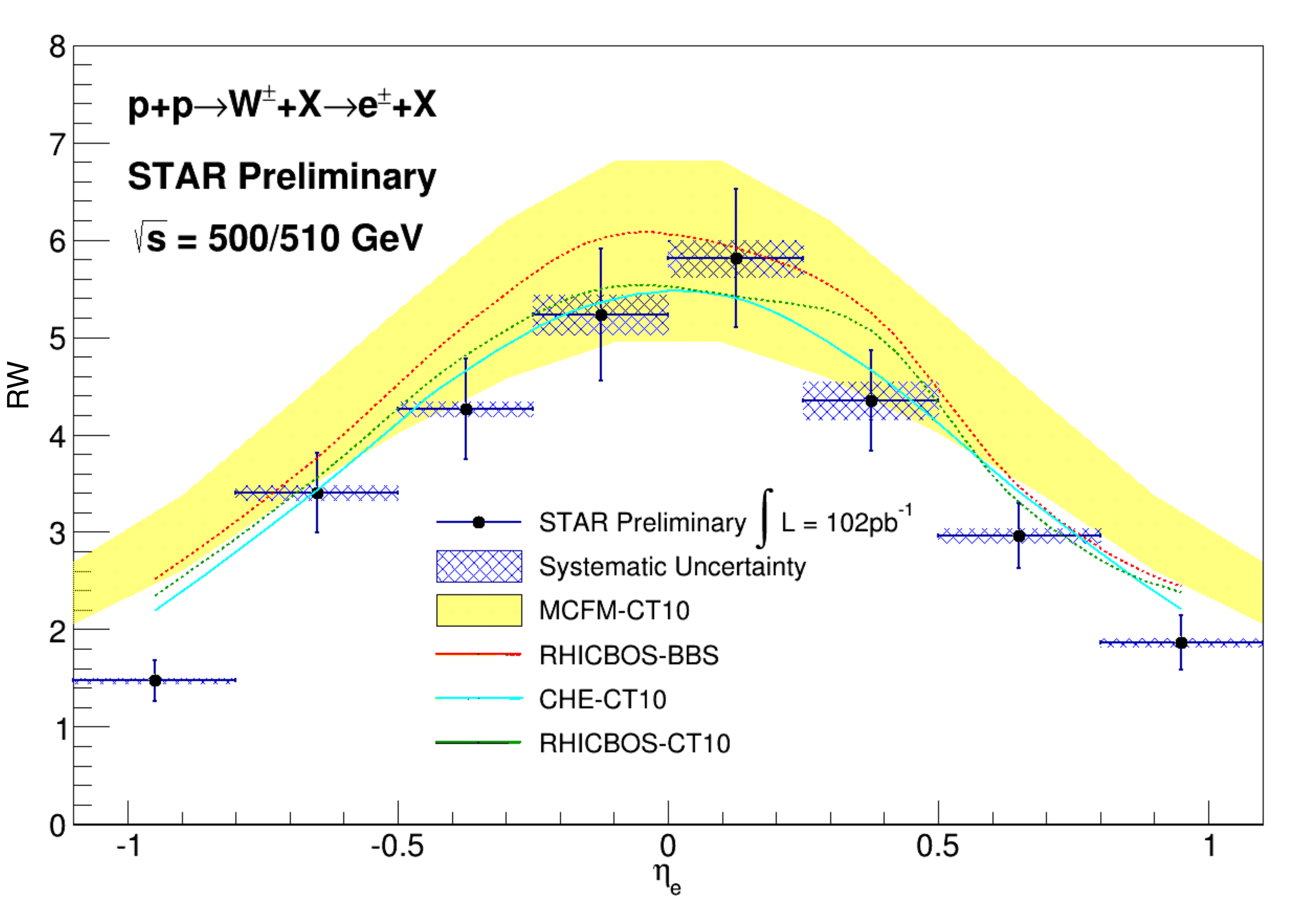}
\caption{$W^+ / W^-$ cross section ratio as a function of electron pseudo-rapidity.}
\label{Fig:results2}
\end{figure}

\begin{figure}[!h]
\centering
\includegraphics[width=0.48\textwidth]{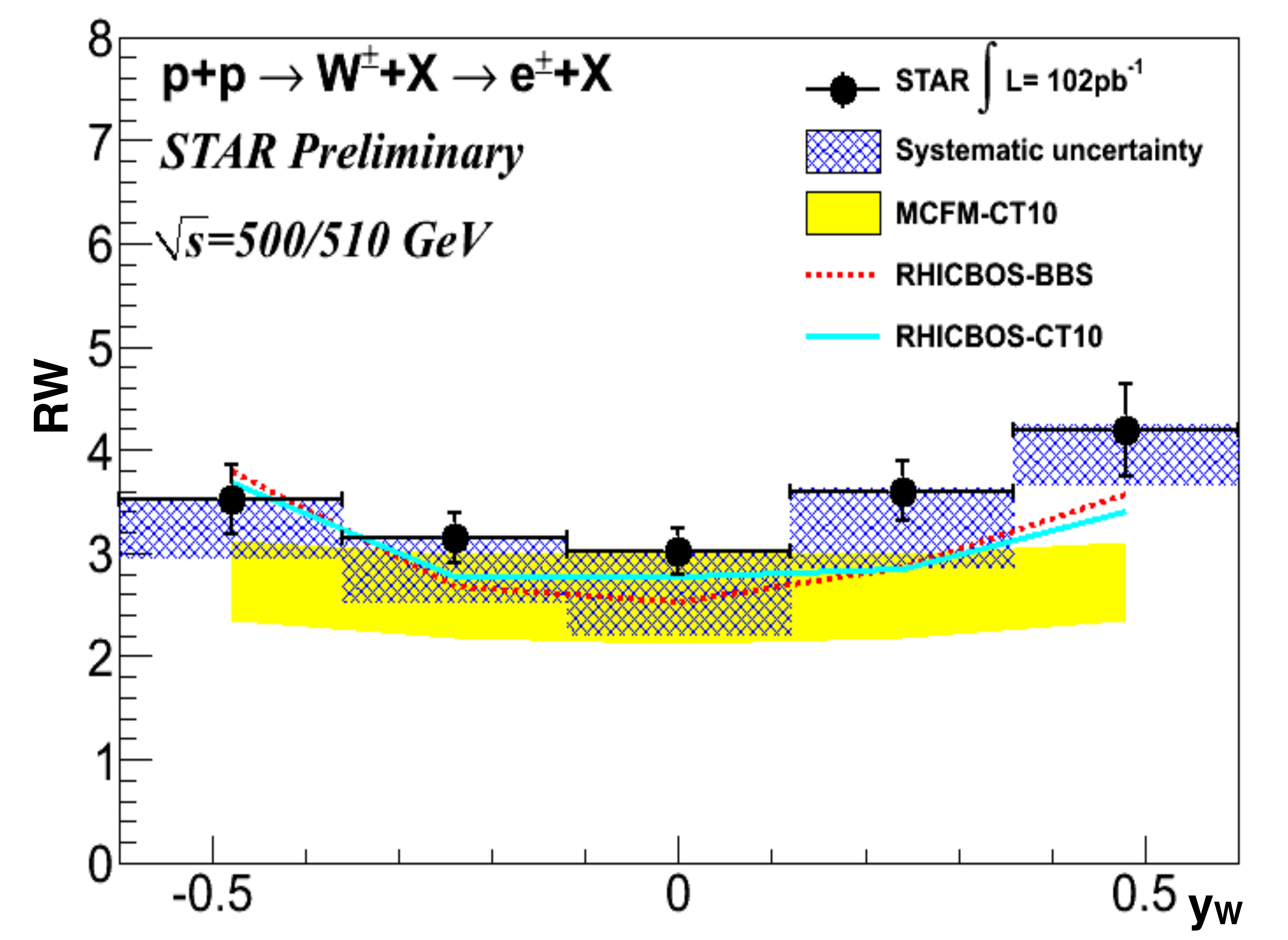}
\caption{$W^+ / W^-$ cross section ratio as a function of W boson rapidity}
\label{Fig:results3}
\end{figure}

The  $W^\pm$ single-spin asymmetries were calculated using the formula as shown in equation~\ref{eq:AL},
\begin{equation}
A^W_L=\frac{A_L - \alpha}{\beta}
\label{eq:AL}
\end{equation}
where $A_L$ is the parity violating single spin asymmetry which is dominated by the signal but still contain some residual background. This raw asymmetry was calculated separately for each RHIC beam for a given $\eta$ bin at STAR in terms of luminosity corrected yields, $N_i$ corresponds to four RHIC helicity states, $+ +,+ -,- + and - -$ and average beam polarization of the beam $P_{1,(2)}$. The polarized background correction $\alpha$, was determined using the fraction of Z background component contained in the signal yield and its respective longitudinal single-spin asymmetry, $A^Z_L$ which was estimated using full next-to-leading (NLO) order framework~\cite{dis:run9paper}. This contribution was found to be negligible within the statistical uncertainty due to the small ($<$1 \%) fraction of Z background contribution as shown in Fig.~\ref{Fig:bg}. The unpolarized background correction, $\beta$ for $A_L$ is calculated independently for $W^+$ and $W^-$ for each $\eta$ bin. This dilution factor is due to background events as shown in Fig.~\ref{Fig:bg} and is determined by $\beta = S / (S+B) $, where $S (B)$ is the number of signal (background) events for $25 < E^e_T < 50$ GeV. \\
\begin{figure}[!h]
\centering
\includegraphics[width=0.49\textwidth]{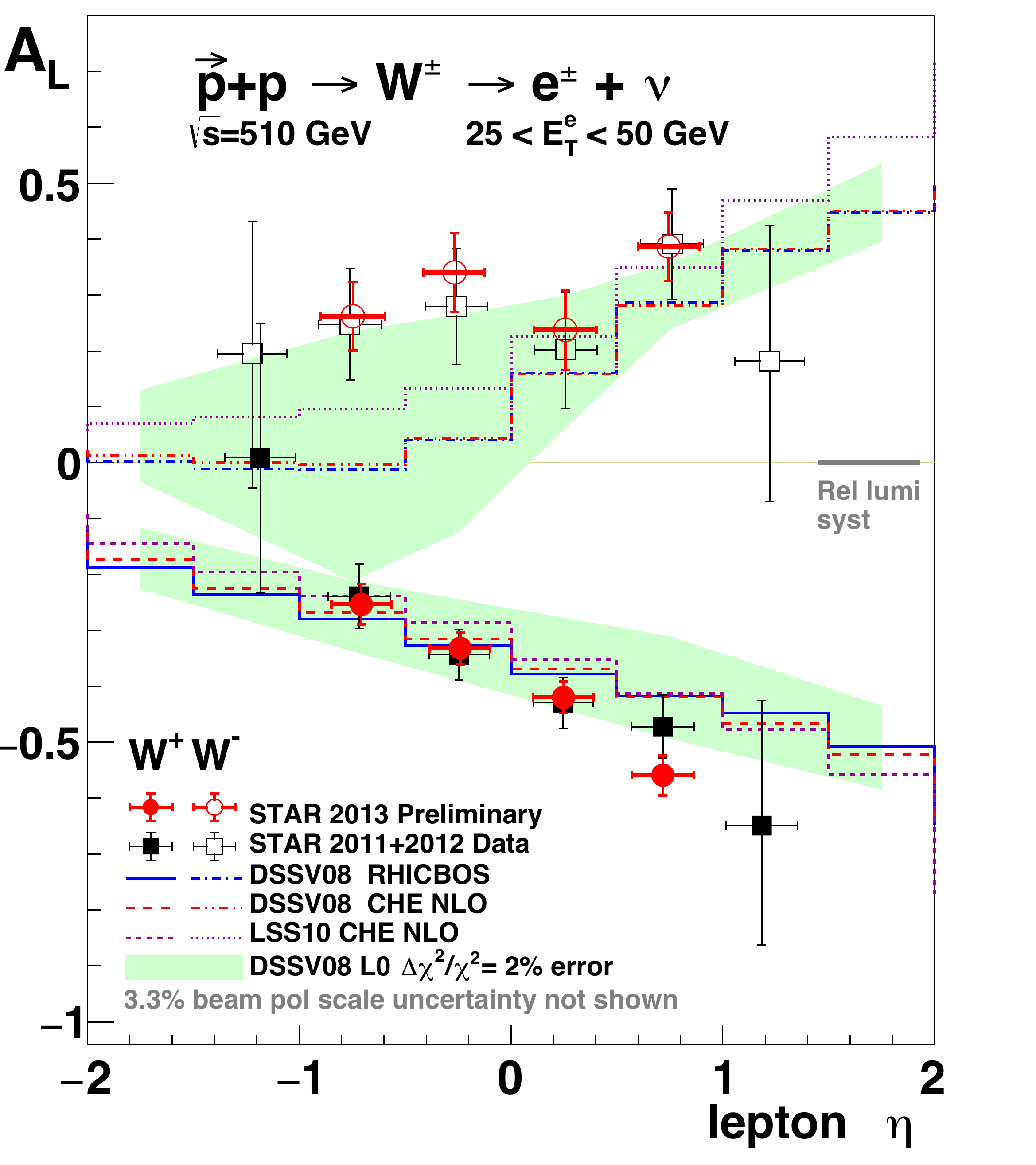}
\caption{Longitudinal single-spin asymmetries for $W^\pm$ production as a function of lepton pseudorapidity, $\eta_e$ in comparison to theory predictions}
\label{Fig:results}
\end{figure}
\indent
The STAR 2012 published~\cite{dis:run12paper} $W^\pm$ single-spin asymmetry results (open and closed black squares) measured for $e^\pm$ are shown in Fig.~\ref{Fig:results} along with recently released STAR 2013 preliminary results (open and closed red circles) as the function of decay $e^\pm$ pseudorapidity, $\eta_e$ in comparison to theoretical predictions based on DSSV08~\cite{dis:dssv08} and LSS10~\cite{dis:lss10} helicity-dependent PDF sets, using both CHE (next-to-leading order)~\cite{dis:rhicW}  and RHICBOS (fully resummed) frameworks~\cite{dis:rhicbos}. The new 2013 preliminary results are consistent with published 2012 results which measured  larger $A^{W^-}_L$ than the central value of the theoretical predictions. The enhancement at large negative $\eta_e$, in particular is sensitive to the polarized anti u quark distribution, $\Delta\bar u$. $A^{W^+}_L$ is negative as expected and consistent with theoretical predictions. The total uncertainties in both results are completely statistically driven while systematic uncertainties are well under control. Vertical error bars include statistical uncertainty as well as systematic uncertainties due to the unpolarized background dilutions ($<$10 \% of statistical uncertainty). Horizontal error bars of each point represented the width of the $\eta_e$ distributions within the bin. In 2013 results, systematic uncertainty due to the BEMC calibration is represent by the thickness of the horizontal bar which is in the same order of the relative luminosity systematic, represent by the grey band. The total systematic of 2013 results are on the same order of published 2012 results despite significant enhancement in luminosity in RHIC 2013 running in comparison to previous years. The uncertainty of new 2013 preliminary results is reduced by 40\% in comparison to published results making new results the most precise measurement in the world to date. The STAR 2012 preliminary $A^{W^\pm}_L$ results~\cite{dis:global} are included in the DSSV++ global analysis~\cite{dis:dssv++} from the DSSV group and recent NNPDF~\cite{dis:nnpdf} global analysis. Both analyses show that the STAR W $A_L$ results provide a significant constraint on $\Delta\bar u$ and $\Delta\bar d$ quark polarizations. We expect new STAR 2013 preliminary results to further constrain anti u ($\Delta\bar u$) and anti d ($\Delta\bar d$) quark polarizations. \\
\indent
The charged W cross section ratio can be measured experimentally as 
\begin{equation}
\frac{\sigma _{W^+}}{\sigma _{W^-}} = \frac{N^+_i - N^+_B }{N^-_i - N^-_B}\frac{\varepsilon^-}{\varepsilon^+}
\label{eq:Xsection}
\end{equation}
where $\pm$ corresponds to positively or negatively charged lepton, $N_i$ are reconstructed $W^\pm$ yields  from decay $e^\pm$, $N_B$ are estimated background yields  and $\varepsilon$ is the efficiency at which W events are detected. The detection efficiencies which account for all cut and detector efficiencies  are calculated using Monte Carlo based on PYTHIA 6.422 and GEANT simulations. However there was only a small ($\sim 1-2\%$) charge dependence measured between the $W^+$ and $W^-$ efficiencies leading to a negligible contribution to the charged W cross section ratio. Figure~\ref{Fig:results2}(\ref{Fig:results3}) shows the charged W cross section ratio for the combined 2011 and 2012 runs, computed using equation~\ref{eq:Xsection} as a function of the electron pseudo-rapidity $\eta_e$ (W boson rapidity, $y_W$). More information on how the W boson kinematics were reconstructed can be found in \cite{dis:wrapidity}. 
The error bars on data points represent the statistical uncertainty, while the shaded boxes correspond to the systematic uncertainty. The yellow band and colored curves serve as a comparison to different PDF sets \cite{dis:XsecPDF1,dis:XsecPDF2} and theory frameworks \cite{dis:XsecTheory1,dis:XsecTheory2}. The systematic uncertainties for the charged W cross section ratios as a function of $\eta_e$ are due to the background subtraction~\cite{dis:matt} and are well under control similar to the asymmetry analysis. The systematic uncertainty of the result as a function of $y_W$ is contributed both by the background subtraction and W reconstruction smearing~\cite{dis:wrapidity} where the latter provided the leading contribution. Further studies into this newly established W boson kinematics reconstruction process~\cite{dis:wrapidity} should reduce the systematic uncertainties on the $W^\pm$ cross-section ratio dependence on the boson kinematics. 
\section{IV. SUMMARY AND OUTLOOK}
We present the STAR 2013 preliminary results of measurements of single spin asymmetries for $W^\pm$ boson production in longitudinally polarized p+p collisions and STAR 2011+2012 W cross section ratios $W^+ / W^-$, at $\sqrt{s}=510$ GeV. The new 2013  $A^{W^-}_L$  results are consistent with STAR published 2012 $A^{W^-}_L$ results, further confirming the measured large $W^-$ asymmetry compared to the theoretical prediction indicating a large anti u quark polarization. Furthermore, the uncertainty of new 2013 results is reduced by 40\% in comparison to published results making the STAR 2013 preliminary results the most precise measurements of $A^{W^-}_L$  in the world up to date. The uncertainties are purely statistical driven while systematics are well under control. With the reduced uncertainty, we expect our new results to further constrain the antiquark helicity distribution functions. Analysis is ongoing to measure the asymmetry in the forward rapidity region from  $1<\eta<1.4$ using STAR EEMC subsystem and further extend using STAR Forward Gem Tracker (FGT) which covers the acceptance between $1<\eta<2$. This enhances the sensitivity to $\bar u$ and $\bar d$ quark polarizations. We expect to include these measurements in the publication from STAR 2013 data. STAR has also  measured and presented charged W cross section ratios from combined 2011 and 2012 proton-proton STAR data at $\sqrt{s}$ = 500 and 510 GeV. The inclusion of this data into global PDF analysis should help constrain the sea quark distributions and provide additional insight into the $\bar d / \bar u$ ratio at relatively higher Bjorken-x values where the behavior of $\bar d / \bar u$ is not clearly understand yet.



\begin{thebibliography}{99}

\bibitem{dis:rqpModel}
J. Kuti and V. F. Weisskopf, Phys. Rev. {\bf{D4}}, 3418 (1971).


\bibitem{dis:EMU}
European Muon Collaboration, J. Ashman \emph{et al}., Phys. Rev. Lett. {\bf{B206}}, 364 (1988).

\bibitem{dis:sumrule}
R. Jaffe and A. Manohar, Nucl. Phys. {\bf{B337}}, 509 (1990), revised version.

\bibitem{dis:global} 
D. de Florian, R. Sassor, M. Stratmann, and W. Vogelsang, Phys. Rev. {\bf{D80}}, 034030 (2009).

\bibitem{dis:frag}
B. Adeva \emph{et al}., (Spin Muon Collaboration), Phys. Lett.{\bf{B420}}, 180 (1998).

\bibitem{dis:newdata1}
A. Airapetian et al. (HERMES Collaboration), Phys. Rev. {\bf{D71}}, 012003 (2005).

\bibitem{dis:newdata2}
M. G. Alekseev et al. [COMPASS Collaboration], Phys. Lett. {\bf{B680}}, 217 (2009).

\bibitem{dis:newdata3}
M. G. Alekseev et al. [COMPASS Collaboration], Phys. Lett. {\bf{B693}}, 227 (2010).

\bibitem{dis:FFs}
D. de Florian, R. Sassot, and M. Stratmann, Phys. Rev. {\bf{D75}}, 114010 (2007); {\bf{D76}}, 074033 (2007).

\bibitem{dis:lss10}
E. Leader, A. V. Sidorov, and D. B. Stamenov, Phys. Rev. {\bf{D82}}, 114018 (2010).

\bibitem{dis:rhicW} 
D. de Florian and W. Vogelsang, Phys. Rev. {\bf{D81}}, 094020 (2010).

\bibitem{dis:GSR}
R. D. Field and R. P. Feynman, Phys. Rev. D15 (1977) 2590.

\bibitem{dis:E866}
R. S. Towell et al., Phys. Rev. {\bf{D64}}, 052002 (2001).

\bibitem{dis:FAtheory}
G.T. Garvey and J.C. Peng, Progress in Particle and Nuclear Physics, {\bf{47}}, 203-243 (2001)

\bibitem{dis:seaquest}
B. Kerns et al. (SeaQuest Collaboration), APS April Meeting, 2016

\bibitem{run11paper}
STAR Collaboration et al.(2015)]{2015arXiv151106003S} STAR Collaboration, Adamczyk, L., Adkins, J.~K., et al.\ 2015, arXiv:1511.06003 

\bibitem{dis:starNIM}
K.H. Ackermann {et al}., Nucl. Instrum. Meth. {\bf{A 499}}, 624 (2003)

\bibitem{dis:run12paper} 
STAR Collaboration, L. Adamczyk \emph{et al}., Phys. Rev. Lett. {\bf{113}}, 072301 (2014).

\bibitem{dis:antikt}
M. Cacciari, G. P. Salam, and G. Soyez, JHEP {\bf{0804}}, 063 (2008).

\bibitem{dis:pythia}
T. Sjostrand, S.Mrenna, and P.Z. Skands, JHEP {\bf{05}}, 026 (2006).

\bibitem{dis:geant}
R. Brun \emph{et al}., CERN-DD-78-2-REV (1978).

\bibitem{dis:run9paper} 
M.~M.~Aggarwal {\it et al.} [STAR Collaboration], Phys.\ Rev.\ Lett.\  {\bf 106}, 062002 (2011), and references therein.
  
\bibitem{dis:dssv08}
D. de Florian, R. Sassor, M. Stratmann, and W. Vogelsang, Phys. Rev. Lett. {\bf{101}}, 072001 (2008).

\bibitem{dis:rhicbos}
P. M. Nadolsky and C. Yuan, Nucl. Phys. {\bf{B666}}, 31 (2003)

\bibitem{dis:dssv++}
E. Aschenauer \emph{et al}., (2013), arXiv:1304.0079 [nucl-ex].

\bibitem{dis:nnpdf}
E.R. Nocera, arXiv:1403:0440 [hep-ph] (2014).

\bibitem{dis:wrapidity}
 S. Fazio and D. Smirnov (STAR), PoS, DIS2014, 237, (2014)

\bibitem{dis:XsecPDF1}
H. L. Lai et al., Phys. Rev. {\bf{D82}}, 074024 (2010)

\bibitem{dis:XsecPDF2}
C. Bourrely, F. Buccella, and J. Soffer, Eur. Phys. J. C 23, 487 (2002)

\bibitem{dis:XsecTheory1}
J. Campbell, K. Ellis, and C. Williams, \textit{MCFM - Monte Carlo for FeMtobarn Processes}, mcfm.fnal.gov.

\bibitem{dis:XsecTheory2}
 P .M. Nadolsky and C.-P. Yuan, Nucl. Phys. {\bf{B666}}, 3 (2003)

\bibitem{dis:matt}
M. Posik  (STAR), Constraining Sea Quark Distributions Through $W^{\pm}$ Cross Section Ratios Measured at STAR, DIS2015, arXiv 1507.07854

\end{thebibliography}

\end{document}